\documentclass[11pt]{article}
\usepackage{amsmath}
\usepackage{amsthm}
\usepackage{titlesec}
\usepackage{indentfirst}
\theoremstyle{definition}\newtheorem{Df}{Definition}
\theoremstyle{plain}\newtheorem{Th}{Theorem}
\theoremstyle{definition}\newtheorem{Rm}{Remark}
\theoremstyle{definition}
\theoremstyle{plain}\newtheorem{Pp}[Th]{Proposition}
\theoremstyle{plain}
\theoremstyle{plain}\newtheorem{Lm}[Th]{Lemma} \textwidth 155mm
\begin{document}
\title{{\bf Determining  the equivalence for 1-way quantum finite automata\thanks{This
research is supported  by the National Natural Science Foundation
(Nos. 90303024, 60573006), the Research Foundation for the
Doctorial Program of Higher School of Ministry of Education (No.
20050558015), and the Natural Science Foundation of Guangdong
Province (No. 031541) of China.}}}
\author{Lvzhou Li,\hskip 2mm Daowen Qiu\thanks{Corresponding author.  {\it E-mail
address:}
issqdw@mail.sysu.edu.cn (D. Qiu).} \\
\small{{\it Department of
Computer Science, Zhongshan University, Guangzhou 510275,}}\\
\small {{\it People's Republic of China}}}
\date{ }
\maketitle \vskip 2mm \noindent {\bf Abstract}
\par
Two quantum finite automata are equivalent if for any input string
$x$ the two automata accept $x$ with equal probability.  In this
paper, we focus on  determining the equivalence for  {\it 1-way
quantum finite automata with control language} (CL-1QFAs) defined
by Bertoni et al
 and {\it measure-many 1-way quantum finite automata}
(MM-1QFAs) introduced by Kondacs and Watrous. It is  worth
pointing out that although Koshiba
 tried to solve the same problem for MM-1QFAs, we show
that his method  is not valid, and thus determining the
equivalence between MM-1QFAs is still left open until this paper
appears. More specifically, we obtain that:
\begin{enumerate}
    \item[(i)] Two CL-1QFAs ${\cal A}_1$ and ${\cal A}_2$ with control
    languages (regular languages)
${\cal L}_1$ and  ${\cal L}_2$, respectively, are equivalent if
and only if they are $(c_1n_1^2+c_2n_2^2-1)$-equivalent, where
$n_1$ and $n_2$ are the numbers of states in ${\cal A}_1$ and
${\cal A}_2$, respectively, and $c_1$ and $c_2$ are the numbers of
states in the minimal DFAs that recognize ${\cal L}_1$ and  ${\cal
L}_2$, respectively. Furthermore, if ${\cal L}_1$ and ${\cal L}_2$
are given in the form of DFAs, with $m_1$ and $m_2$ states,
respectively, then there exists a polynomial-time algorithm
running in time $O ((m_1n_1^2+m_2n_2^2)^4)$ that takes as input
${\cal A}_1$ and ${\cal A}_2$ and determines whether they are
equivalent.
    \item[(ii)]  Two MM-1QFAs ${\cal A}_1$ and ${\cal A}_2$  with $n_1$ and $n_2$
states, respectively, are equivalent if and only if they are
$(3n_1^2+3n_2^2-1)$-equivalent. Furthermore, there is a
polynomial-time algorithm running in time $O ((3n_1^2+3n_2^2)^4)$
that takes as input ${\cal A}_1$ and ${\cal A}_2$
 and determines whether ${\cal A}_1$ and ${\cal A}_2$ are
equivalent.
\end{enumerate}


\vskip 2mm

\section{Introduction}
Over the past two decades, quantum computing has attracted wide
attention in the academic community \cite{Gru99,Nie}. To a certain
extent, this was motivated by the exponential speed-up of Shor's
quantum algorithm for factoring integers in polynomial time
\cite{Shor} and afterwards Grover's algorithm of searching in
database of size $n$ with only $O(\sqrt{n})$ accesses
\cite{Grover}. As we know, these algorithms are based on {\it
quantum Turing machines} or {\it quantum circuits} that seem to be
complicated to implement using today's experiment technology.
Therefore, it is natural to consider the simpler models of quantum
computation.

 Classically, {\it finite automata} (FAs), as one of the
 simplest models of computattion, have been deeply studied
\cite{Hop}. Then, as a quantum variant of FAs, {\it quantum finite
automata} (QFAs) are developed and have received  extensive
attention from the academic community. QFAs were first introduced
independently by Moore and Crutchfield \cite{Moore}, as well as
Kondacs and Watrous \cite{Kon97}, and then they were intensively
investigated by others [2-12]. QFAs can be mainly divided into two
kinds: {\it one-way quantum finite automata} (1QFAs) whose tape
heads only move one cell to right at each evolution, and  {\it
two-way quantum finite automata} (2QFAs), in which the tape heads
are allowed to move towards right or left, or to be stationary.
(Notably, Amano and Iwama \cite{Ama} dealt with an intermediate
form called 1.5QFAs, whose tape heads are allowed to move right or
to be stationary, and, particularly, they showed that the
emptiness problem for this restricted model is undecidable.)

Furthermore, by means of the measurement times in a computation,
1QFAs have two fashions: {\it measure-once} 1QFAs (MO-1QFAs)
proposed by Moore and Crutchfield \cite{Moore}, and, {\it
measure-many} 1QFAs (MM-1QFAs) studied first by Kondacs and
Watrous \cite{Kon97}. In addition, in terms of the kind of
measurement allowed, both MO-1QFAs and MM-1QFAs   allow only very
restricted measurement: MO-1QFAs allow  projective measurement
with only two results: acceptance and rejection; MM-1QFAs allow
 projective measurement with only three results: acceptance,
rejection and continuation. As we know, measurement is an
important operation in quantum computation and quantum
information. Then in Refs. \cite{Nay, Ber03,Amb06}, some more
general quantum models were proposed and characterized, in which
any projective measurement was allowed as a valid intermediate
computational step. Particularly, Bertoni et al \cite{Ber03}
characterized a model called {\it 1-way quantum finite automata
with control language} (CL-1QFAs). We will detail it later on.

 In addition to QFAs, there are some
other types of finite-like quantum automata that are being
developed, such as {\it quantum push-down automata} (QPDAs)
\cite{Gol}, {\it quantum one-counter automata} \cite{Yam}, and
{\it quantum sequential machines} (QSMs) \cite{Gud, Qiu}. Some
interesting results have been obtained on these models, and we do
not detail them here.

So far,  work on QFAs  mainly focuses on characterizing the
language recognized by QFAs and comparing them with  their
classical analogies (finite automata and {\it probabilistic finite
automata} \cite{Rab, Paz}). We briefly state some main results
obtained. The class of languages recognized by CL-1QFAs with
bounded error probabilities is strictly bigger than that by
MM-1QFAs which, in turn, recognize the class of languages strictly
bigger than that by MO-1QFAs. However, all of them recognize only
 subclass of regular languages with bounded error
probabilities \cite{Moore,Kon97,Bro,Ber03}. Also, the class of
languages recognized by MM-1QFAs with bounded error probabilities
is not closed under the binary Boolean operations
\cite{Bro,Ber03}. Concerning 2QFAs, an exciting result was
obtained by Kondacs and Watrous \cite{Kon97}  that some 2QFA can
recognize non-regular language $L_{eq}=\{a^{n}b^{n}|n>0\}$ with
one-sided error probability in linear time, which can not be
attained by the classical analogies (even by {\it  two-way
probabilistic automata}).

Although more and more problems concerning  the models of quantum
computation have been clarified, there are still  some fundamental
problems left open. One of these  problems is to determine the
equivalence for these models. As we know, determining the
equivalence for computing models is a very important issue in the
theory of classical computation. For instance, \cite{Hop, Paz}
were all devoted  to this issue and good results were obtained.
Concerning the problem of determining the equivalence for QFAs,
there exists only a little work \cite{Bro,Kos} that deals with the
simplest case---MO-1QFAs. Although in \cite{Kos}, Koshiba tried to
solve the problem for MM-1QFAs, we will show   that his method is
not valid, and thus, in fact, the problem for MM-1QFAs is still
left open. To our knowledge, there seems to be no more related
work on this problem, except for some work  on QSMs \cite{Li}.

In this paper, we focus on determining the equivalence between
CL-1QFAs and between  MM-1QFAs.  Sufficient and necessary
conditions for deciding the equivalence are obtained. Also, we
present some polynomial-time algorithms to judge  the equivalence
for CL-1QFAs and MM-QFAs, respectively. The remainder of this
paper is organized as follows.  Some models and related
definitions are introduced in Section 2. Section 3 is the main
part of the paper which is to deal with the problem stated above.
In Subsection 3.1, we introduce some definitions and related
results that will be used in the later subsections. Then we deal
with the equivalence problems  for CL-1QFAs and MM-1QFAs in
Subsection 3.2 and Subsection 3.3, respectively.  Finally, some
conclusion remarks are made in Section 4.

\section{Preliminaries}
\subsection{ Some notation on linear algebra  and quantum mechanics}

 As usual, for non-empty set
$\Sigma$, by $\Sigma^{*}$ we mean the set of all finite length
strings over $\Sigma$, and by $\Sigma^n$ we mean the set of all
strings over $\Sigma$ with length $n$. For $u\in \Sigma^{*}$,
$|u|$ denotes the length of $u$; for example, if
$u=x_{1}x_{2}\ldots x_{m}\in \Sigma^{*}$ where $x_{i}\in \Sigma$,
then $|u|=m$.  For set $S$,  $|S|$ denotes the cardinality of $S$.

Let ${\bf C}$ denote the set of all complex numbers, ${\bf R}$
 the set of all real numbers, and ${\bf C}^{n\times m}$
 the set of $n\times m$ matrices having entries in ${\bf
C}$. Given two matrices $A\in {\bf C}^{n\times m}$ and $B\in{\bf
C}^{p\times q}$, their {\it Kronecker product} is the $np\times
mq$ matrix, defined as \[A\otimes B=\begin{bmatrix}
  A_{11}B & \dots & A_{1m}B \\
  \vdots & \ddots & \vdots \\
  A_{n1}B &\dots & A_{nm}B \
\end{bmatrix}.\] When operations can be performed, we get $(A\otimes B)(C\otimes D)=AC\otimes BD$.
  Matrix $M\in{\bf C}^{n\times n}$ is said to be {\it
unitary} if $MM^\dagger=M^\dagger M=I$, where
 $\dagger$ denotes conjugate-transpose operation. $M$ is said to be {\it
Hermitian} if $M=M^\dagger$. For $n$-dimensional row vector
$x=(x_1,\dots, x_n)$, its norm denoted by $||x||$ is defined as
$||x||=\big(\sum_{i=1}^n x_ix_i^{*}\big)^{\frac{1}{2}}$, where
symbol $*$ denotes conjugate operation.  Unitary matrices preserve
the norm, i.e., $||xM||=||x||$ for each  $x\in {\bf C}^{1\times
n}$ and unitary matrix $M\in{\bf C}^{n\times n}$.  An
$n$-dimensional row vector ${\bf a}=(a_{1}\hskip 1mm a_{2} \dots
a_{n})$ is called {\it stochastic} if $a_{i}\geq 0$
$(i=1,2,\ldots,n)$, and $\sum_{i=1}^{n}a_{i}=1$; in particular,
${\bf a}$ is called a {\it degenerate stochastic vector} if one of
the entries is 1 and the others 0s. A matrix is called {\it
stochastic} if its each row is a stochastic vector.

We would refer the reader to \cite{Gru99,Nie} for a through
treatment on the postulates of quantum mechanics, and here we just
briefly introduce some notation to be used in this paper. For a
quantum system with a finite basic state set $Q=\{q_1,\dots,
q_n\}$, every basic state $q_i$ can be represented by an
$n$-dimensional row vector $\langle q_i|=(0\dots1\dots0)$ having
only 1 at the $i$th entry. At any time, the state of this system
is a {\it superposition} of these basic states and can be
represented by a row vector $\langle \phi|=\sum_{i=1}^nc_i\langle
q_i|$ with $c_i\in{\bf C}$ and $\sum_{i=1}^n|c_i|^2=1$.  If we
want to get some information from a quantum system, then we should
make a measurement on it. Here we consider {\it projective
measurement} (Von Neumann measurement). A projective measurement
is described by an {\it observable}
 that is  a Hermitian matrix ${\cal O}=c_1P_1+\dots +c_s
 P_s$, where $c_i$ is its eigenvalue and, $P_i$ is the projector
 onto the eigenspace corresponding to $c_i$. In this case, the
 projective measurement of ${\cal O}$ has result set $\{c_i\}$ and
 projector set $\{P_i\}$.

 We assume that the operations of addition and multiplication on
 two complex numbers can all be done in constant time, which will
 be used in Section 3 when we analyze the time complexity of the
 algorithms determining the equivalence between QFAs.
\subsection{ Classical computing models}
Firstly, we give a  mathematical model which  is not an actual
computing model but generalizes many classical computing models,
 which will play a foundational role in this paper.
\begin{Df}
 A {\it bilinear  machine}
(BLM)  over the alphabet $\Sigma$ is a four-tuple ${\cal M}=( S,
\pi,\{M(\sigma)\}_{\sigma\in\Sigma},\eta)$, where $S$ is a finite
state set with $|S|=n$, $\pi\in {\bf C}^{1\times n}$, $\eta\in
{\bf C}^{n\times 1}$ and $M(\sigma)\in{\bf C}^{n\times n}$ for
$\sigma\in \Sigma$.
\end{Df}
Associated to  a BLM ${\cal M}$, the {\it word function} $f_{\cal
M}:\Sigma^{*}\rightarrow {\bf C}$ is defined in the way: $f_{\cal
M}(w)=\pi M(w_1)\dots M(w_n)\eta$, where  $w=w_1\dots w_n\in
\Sigma^{*}$. In particular, when $f_{\cal M}(w)\in {\bf R}$ for
every $w\in \Sigma^{*}$,
 ${\cal M}$ is called a {\it real-valued bilinear  machine} (RBLM).

 Turakainen \cite{Tur} defined a model called {\it generalized
automata} (GAs) and characterized the languages recognized by
them. In fact, a GA ${\cal M}$ is just a BLM with the restriction
that $\pi, \eta$, and $M(\sigma)$ $(\sigma\in\Sigma)$ have
components in  ${\bf R}$. The word function $f_{\cal M}$
associated to GA ${\cal M}$ is defined as in the case of BLMs.

  Another important computing model is the so-called  {\it  probabilistic
automata} (PAs) \cite{Rab,Paz}. A PA is a GA with the restriction
that $\pi$ is a stochastic vector, $\eta$ consists of the entries
with 0's and 1's only, and the matrices $M(\sigma)$ $(\sigma\in
\Sigma)$ are stochastic. Then, the word function $f_{\cal M}$
associated to PA ${\cal M}$ has range
 $[0,1]$.

Given a PA ${\cal M}$, if $\pi$ is a degenerate stochastic vector,
and stochastic matrices $M(\sigma)$ $(\sigma\in \Sigma)$ consist
of the entries with 0's and 1's only, then ${\cal M}$ is called a
{\it determine finite automaton} (DFA) \cite{Hop}. Then, the word
function $f_{\cal M}$ associated to DFA ${\cal M}$ has range
$\{0,1\}$. The language ${\cal L}$ recognized by DFA ${\cal M}$ is
defined by the following set:
\begin{align}
{\cal L}=\{w:w\in\Sigma^{*}\text{~and~} f_{\cal M}(w)=1\}.
\end{align}
In this case, we also call the function $f_{\cal M}$ as the {\it
characterization function} of ${\cal L}$, denoted by $\chi_{\cal
L}$, where for any $x\in \Sigma^{*}$, $\chi_{\cal
L}(x)=\begin{cases}1& x\in{\cal
L},\\
0&x\notin{\cal L}.\end{cases}$ It is well known that DFAs can
recognize only regular languages,  and for every regular language
${\cal L}$, there is a {\it minimal } DFA recognizing it.

>From the definitions above, it is readily seen that:
\begin{align*}
DFAs\subset PAs \subset GAs\subset RBLMs\subset BLMs.
\end{align*}

\subsection{ Quantum computing models}
In this paper, only one-way quantum computing models are
considered. So in the sequel, when introducing quantum models, we
always leave out the word ``one-way''.
\paragraph {Measure-Once Quantum Finite Automata (MO-1QFAs)}MO-1QFAs are the simplest
quantum computing models. In this model, the transformation on any
symbol in the input alphabet is realized by a unitary operator. A
unique measurement is performed at the end of a computation.

 More formally, an MO-1QFA with $n$ states and
the input alphabet $\Sigma$ is a four-tuple ${\cal M}=(Q, \pi,
\{U(\sigma)\}_{\sigma\in \Sigma},{\cal O})$, where
\begin{itemize}
\item  $Q=\{q_1,\dots,q_n\}$ is the basic state set; at any time,
the state of ${\cal M}$ is a superposition of these basic states;

 \item $\pi\in {\bf C}^{1\times n}$ with $||\pi||=1$ is
the initial vector;

\item for any $\sigma\in \Sigma$, $U(\sigma)\in{\bf C}^{n\times
n}$ is a unitary matrix;

\item ${\cal O}$ is an observable described by the projectors
$P(a)$ and $P(r)$, with the result set $\{a,r \}$ of which `$a$'
and `$r$' denote ``accept'' and ``reject'', respectively.
\end{itemize}

Given an MO-1QFA ${\cal M}$ and an input word $x_1\dots
x_n\in\Sigma^{*}$, then starting from $\pi$, $U(x_1),\dots,
U(x_n)$ are applied in succession, and at the end of the word, a
measurement of ${\cal O}$ is performed with the result that ${\cal
M}$ collapses into accepting states or rejecting states with
corresponding probabilities. Hence,  ${\cal M}$ defines a word
function $f_{\cal M}$: $\Sigma^{*}\rightarrow [0,1]$ in the
following:
\begin{align}
f_{\cal M}(x_1\dots x_n)=||\pi\Big(\prod^n_{i=1}U(x_i)\Big
)P(a)||^2.\label{f_MO}
\end{align}
For any input word $w\in \Sigma^{*}$, $f_{\cal M}(w)$ denotes the
probability of ${\cal M}$ accepting $w$. Sometimes, we also use
${\cal P}_{\cal M}(w)$ to denote this probability.

\paragraph{Measure-Many Quantum Finite Automata (MM-1QFAs)} Unlike MO-1QFAs that allow only one measurement at the end of a computation,
 MM-1QFAs allow measurement at each step. Due to this difference,
 MM-1QFAs are more powerful than  MO-1QFAs.

Formally,  given an input alphabet $\Sigma$ and an end-maker
$\$\notin\Sigma$, an
 MM-1QFA with $n$ states over the {\it working
 alphabet}
 $\Gamma=\Sigma\cup\{\$\}$ is a four-tuple
 ${\cal M}=(Q,\pi,\{U(\sigma)\}_{\sigma\in\Gamma},{\cal O})$, where
\begin{itemize}
\item  $Q$, $\pi$, and $U(\sigma)$ ($\sigma\in \Gamma$) are
defined as in the case of MO-1QFAs;

\item ${\cal O}$ is an observable  described by the projectors
$P(a)$, $P(r)$ and $P(g)$, with the results in $\{a,r,g \}$ of
which `$a$', `$r$' and `$g$' denote ``accept'', ``reject'' and
``go on'', respectively.
\end{itemize}

Any input word $w$ to MM-1QFAs is in the form: $w\in\Sigma^{*}\$$,
with symbol $\$$ denoting the end of a word. Given an input word
$x_1\dots x_n\$$ where $x_1\dots x_n\in \Sigma^n$, MM-1QFA ${\cal
M}$ performs the following computation:
\begin{itemize}
\item [1.] Starting from $\pi$, $U(x_1)$ is applied, and then we
get a new state $\langle\phi_1|=\pi U(x_1)$. In succession, a
measurement of ${\cal O}$ is performed on $\langle\phi_1|$, and
then the measurement result $i$ ($i\in \{a,g,r\}$) is yielded as
well as  a new state
$\langle\phi_1^{i}|=\frac{\langle\phi_1|P(i)}{\sqrt{p_1^i}}$
 is gotten, with corresponding probability
 $p_1^i=||\langle\phi_1|P(i)||^2$.

 \item [2.]   In the above step, if $\langle\phi_1^{g}|$ is gotten, then  starting from $\langle\phi_1^{g}|$, $U(x_2)$ is
 applied and a measurement of ${\cal O}$ is performed. The
  evolution rule is  the same as the above step.

 \item [3.] The process continues as far as the measurement result
 `$g$' is yielded. As soon as the measurement result is `$a$'(`$r$'), the
 computation halts and the input word is accepted (rejected).
\end{itemize}

Thus, MM-1QFA ${\cal M}$ defines a word function $f_{\cal
M}:\Sigma^{*}\$\rightarrow[0,1]$ in the following:
\begin{align}
f_{\cal M}(x_1\dots
x_n\$)=\sum^{n+1}_{k=1}||\pi\prod^{k-1}_{i=1}\big(U(x_i)P(g)\big)U(x_k)P(a)||^2,
\end{align}
or equivalently,
\begin{align}
f_{\cal M}(x_1\dots
x_n\$)=\sum^n_{k=0}||\pi\prod^k_{i=1}\big(U(x_i)P(g)\big)U(x_{k+1})P(a)||^2
\end{align}
where, for simplicity, we denote $\$$ by $x_{n+1}$ and we will
always use this denotation in the sequel. $f_{\cal M}(x_1\dots
x_n\$)$ is the probability of ${\cal M}$ accepting the word
$x_1\dots x_n$, and usually, we would like to use another function
denoted by ${\cal P}_{\cal M}:\Sigma^{*}\rightarrow[0,1]$ to
denote this probability such that ${\cal P}_{\cal M}(x_1\dots
x_n)=f_{\cal M}(x_1\dots x_n\$)$.

\paragraph{Quantum Automata with Control Language(CL-1QFAs)}  Bertoni et al \cite{Ber03}
introduced a new 1-way quantum computing model that allows a more
general measurement than the previous  models. Similar to the case
in MM-1QFAs, the state of this model can be observed at each step,
but an observable ${\cal O}$ is considered with a fixed, but
arbitrary, set of possible results ${\cal C}=\{c_1,\dots,c_n\}$,
without limit to $\{a,r,g\}$ as in MM-1QFAs. The accepting
behavior in this model is also different from that of the previous
models. On any given input word $x$, the computation displays a
sequence $y\in {\cal C}^{*}$ of results of ${\cal O}$ with a
certain probability $p(y|x)$, and the computation is accepted if
and only if $y$ belongs to a fixed regular control language ${\cal
L}\subseteq {\cal C}^{*}$.

More formally, given an input alphabet $\Sigma$ and the end-marker
  symbol $\$\notin\Sigma$, a  CL-1QFA over the working
  alphabet $\Gamma=\Sigma\cup\{\$\}$ is a five-tuple ${\cal
  M}=(Q, \pi,\{U(\sigma)\}_{\sigma\in\Gamma},{\cal O},{\cal L})$, where
\begin{itemize}
\item $Q$, $\pi$ and $U(\sigma)$ $(\sigma\in\Gamma)$ are defined
as those of the two previous  quantum models;

\item ${\cal O}$ is an observable with the set of possible results
${\cal C}=\{c_1,\dots,c_s\}$ and  the projector set
$\{P(c_i):i=1,\dots,s\}$ of which $P(c_i)$ denotes the projector
onto the eigenspace corresponding to $c_i$;

\item  ${\cal L}\subseteq{\cal C}^{*}$ is a regular language
(control language).
\end{itemize}

The input word $w$ to CL-1QFA ${\cal M}$ is in the form:
$w\in\Sigma^{*}\$$, with symbol $\$$ denoting the end of a word.
Now, we define the behavior of ${\cal M}$ on word $x_1\dots
x_n\$$. The computation starts in the state $\pi$, and then the
transformations associated with symbols in the word  $x_1\dots
x_n\$$ are applied in succession. The transformation associated
with any symbol $\sigma\in\Gamma$ consists of two steps:
\begin{enumerate}
\item[1.] Firstly, $U(\sigma)$ is applied to the current state
$\langle\phi|$ of ${\cal M}$, yielding the new state
$\langle\phi^{'}|=\langle\phi|U(\sigma)$.

\item[2.] Secondly, the observable ${\cal O}$ is measured on
$\langle\phi^{'}|$. According to quantum mechanics principle, this
measurement yields result $c_k$ with probability
$p_k=||\langle\phi^{'}|P(c_k)||^2$, and the state of ${\cal M}$
collapses to $\langle\phi^{'}|P(c_k)/\sqrt{p_k}$.
\end{enumerate}

Thus, the computation on word $x_1\dots x_n\$$ leads to a sequence
$y_1\dots y_{n+1}\in {\cal C}^{*}$ with probability $p(y_1\dots
y_{n+1}|x_1\dots x_n\$)$ given by
\begin{align}
p(y_1\dots y_{n+1}|x_1\dots
x_n\$)=||\pi\prod^{n+1}_{i=1}U(x_i)P(y_i)||^2,
\end{align}
where we let $x_{n+1}=\$$ as stated before. A computation leading
to the word $y\in {\cal C}^{*}$ is said to be  accepted if $y\in
{\cal L}$. Otherwise, it is rejected. Hence, CL-1QFA ${\cal M}$
defines a word function $f_{\cal M}: \Sigma^{*}\$\rightarrow[0,1]$
in the form:
\begin{align}
f_{\cal M}(x_1\dots x_n\$)=\sum_{y_1\dots y_{n+1}\in {\cal
L}}p(y_1\dots y_{n+1}|x_1\dots x_n\$),\label{f_CL}
\end{align}
which denotes the probability of ${\cal M}$ accepting the word
$x_1\dots x_n$. Usually, we also denote this accepting probability
by the function ${\cal P}_{\cal M}: \Sigma^{*}\rightarrow[0,1]$
where
\begin{align}{\cal P}_{\cal M}(x_1\dots x_n)=f_{\cal M}(x_1\dots x_n\$).\end{align}.

\section{Determining the equivalence for  computing models}
Determining the equivalence for computing models is  an important
issue in the theory of computation.  However, this problem has not
been  well investigated for quantum computing models.  In this
section, we will deal with the equivalence problem for CL-1QFAs
and MM-1QFAs. Our idea is to first transform these quantum models
to BLMs, and then we deal with the equivalence problem for BLMs.
So, below we first give some results on BLMs.
\subsection{Some definitions  and   results  on BLMs to be used}
Firstly, for $x_1\dots x_n\$$ where $\$$ denotes the end-marker,
we mean $|x_1\dots x_n\$|=|x_1\dots x_n|=n$. Now we give two
definitions concerning the equivalence for models.
\begin{Df}
Two BLMs (including RBLMs, GAs, PAs,  and DFAs) ${\cal M}_1$ and
${\cal M}_2$ over the same alphabet $\Sigma$ are said to be
equivalent (resp. $k$-equivalent) if $f_{{\cal M}_1}(w)=f_{{\cal
M}_2}(w)$ for any $w\in \Sigma^{*} $ (resp. for any input string
$w$ with $ |w|\leq k$).
\end{Df}
\begin{Df}
Two QFAs (including MO-1QFAs, MM-1QFAs, and CL-1QFAs) ${\cal M}_1$
and ${\cal M}_2$ over the same input alphabet $\Sigma$ are said to
be equivalent (resp. $k$-equivalent) if ${\cal P}_{{\cal
M}_1}(w)={\cal P}_{{\cal M}_2}(w)$ for any $w\in \Sigma^{*} $
(resp. for any input string $w$ with $ |w|\leq k$).
\end{Df}
Next we  give two propositions concerning BLMs that will be used
later. The first one is Proposition 1 in the following that allows
us to remove the end-maker $\$$ in the input word, as we will see
when we deal with the equivalence for QFAs.

\begin{Pp}
 Let BLM ${\cal M}$
have $n$ states and the alphabet $\Sigma\cup\{\tau\}$ where
$\tau\notin\Sigma$. Then we can give another  BLM $\hat{\cal M}$
over the alphabet $\Sigma$ with the same states, such that
$f_{{\cal M}}(w\tau)=f_{\hat{\cal M}}(w)$, for any $w\in
\Sigma^*$.
\end{Pp}
\noindent\textbf{Proof.} We  let $\hat{\cal M}$ be the same as
${\cal M}$ except that ${\hat{\eta}}=U(\tau).\eta$, where
${\hat{\eta}}$ belongs to $\hat{\cal M}$ and $U(\tau)$ and $\eta$
belong to ${\cal M}$. It is clear that $f_{{\cal
M}}(w\tau)=f_{\hat{\cal M}}(w)$, for any $w\in \Sigma^*$.\qed\\

The second one is Proposition 2 that allows us to convert the
problems in the field of complex numbers to the ones in the field
of real numbers. The idea behind this proposition was first
pointed out by Moore and Crutchfield in \cite{Moore}.
\begin{Pp}
 For any RBLM ${\cal M}$ with $n$ states and the alphabet
$\Sigma$, we can construct effectively an equivalent GA ${\cal
M}^{'}$ with $2n$ states and the same alphabet $\Sigma$.
\end{Pp}
\noindent\textbf{Proof.}  It is well known that
 any complex number $c=a+bi$  has a {\it real matrix
 representation } in the form $ c=
\begin{bmatrix}
  a & b \\
  -b & a
\end{bmatrix}
 $. Then in the same way any $n\times n$
 complex matrix has a representation by a $2n\times 2n$ real
 matrix. We can also check that given two matrices $A$ and $B$ (assuming that $A$ and $B$ can multiply), if $A^{'}$ and $B^{'}$ are
 the real matrix representations of $A$ and $B$, respectively, then
 $A^{'}B^{'}$ will be the real matrix representation of $AB$.

 Now suppose that we have an $n$-state RBLM ${\cal M}=( S,
\pi,\{M(\sigma)\}_{\sigma\in\Sigma},\eta)$. Then for $x=x_1\dots
x_m\in \Sigma^{*}$, there is $\pi M(x_1)\dots M(x_m)\eta=f_{\cal
M}(x)\in {\bf R}$. Using the above representation, we transform
$\pi$, $M(x_i)$ and $\eta$ into $2\times 2n$, $2n\times 2n$ and
$2n\times 2$ real matrices $\hat{\pi}$, $\hat{M}(x_i)$ and
$\hat{\eta}$, respectively. Then we have
\begin{align}
\hat{\pi}\hat{M}(x_1)\dots\hat{M}(x_m)\hat{\eta}=
\begin{bmatrix}
  f_{\cal M}(x) & 0 \\
  -0 & f_{\cal M}(x)
\end{bmatrix}.
\end{align}
Letting $\pi^{'}$ be the top row of $\hat{\pi}$ and $\eta^{'}$ the
left column of $\hat{\eta}$, and letting
$M^{'}(\sigma)=\hat{M}(\sigma)$ for $\sigma\in\Sigma$, we get the
expected GA ${\cal M}^{'}=(S^{'},\pi^{'},
\{M^{'}(\sigma)\}_{\sigma\in\Sigma}, \eta^{'})$, such that
$f_{\cal M}(w)=f_{{\cal M}^{'}}(w)$ for any $w\in \Sigma^{*} $.
Therefore, we have completed the proof. \qed\\

 PAs, as a special case of BLMs, have been well
studied. Specially, concerning the equivalence between PAs, Paz
\cite{Paz} obtained an important result  as follows.
\begin{Th}[{\cite{Paz}}]
Two PAs ${\cal A}_1$ and ${\cal A}_2$ with $n_1$ and $n_2$ states,
respectively, are equivalent if and only if they are
$(n_1+n_2-1)$-equivalent.
\end{Th}
Although Theorem 3 provides a necessary and sufficient condition
for the equivalence between PAs, directly testing it  needs
exponential time. Then Tzeng \cite{Tze} further provided a
polynomial-time algorithm to determine whether two PAs are
equivalent. Hence, the equivalence problem of  PAs has been solved
completely.
\begin{Th}[{\cite{Tze}}]
  There is a polynomial-time algorithm running in time $O((n_1+n_2)^4)$
that takes as input two PAs ${\cal A}_1$ and ${\cal A}_2$ and
determines whether ${\cal A}_1$ and ${\cal A}_2$ are equivalent,
where $n_1$ and $n_2$ are the numbers of states in ${\cal A}_1$
and ${\cal A}_2$, respectively.
\end{Th}

In fact, if we refer to \cite{Paz,Tze} and read carefully the
proofs of Theorem 3 and Theorem 4, then we can  find that their
proofs did not use any essential property of PAs,  just based on
some ordinary knowledge on matrix and linear space, and as a
result, the proofs can also be extended to BLMs. Thus, we get a
more general result as follows.
\begin{Pp}
Two BLMs (including RBLMs, GAs,  PAs, and DFAs) ${\cal A}_1$ and
${\cal A}_2$ with $n_1$ and $n_2$ states, respectively, are
equivalent if and only if they are $(n_1+n_2-1)$-equivalent.
Furthermore, there exists a polynomial-time algorithm running in
time $O((n_1+n_2)^4)$ that takes as input two BLMs ${\cal A}_1$
and ${\cal A}_2$ and determines whether ${\cal A}_1$ and ${\cal
A}_2$ are equivalent.
\end{Pp}
\begin{Rm}
(i) The algorithm for BLMs performs  the same process as that for
PAs.
 The  consuming time of the algorithms for them may differ by a
 constant factor, but with the same magnitude $O((n_1+n_2)^4)$,
 because that  BLMs are considered in the field of complex
 numbers while PAs are restricted to the field of real numbers.
 (ii) When
designing the algorithm for RBLMs, in order to avoid the operation
on complex numbers, one may first transform  RBLMs to  GAs by
Proposition 2, and then determine the equivalence for GAs, using
the algorithm stated in \cite{Tze}. However, the transforming
process is not necessary.
\end{Rm}

Now we turn to the problem of determining the equivalence for
1QFAs. For the equivalence between MO-1QFAs, some solutions have
been obtained by
 \cite{Bro,Kos}. Their idea is   to firstly  transform MO-1QFAs to RBLMs  by the bilinearization technique stated
in \cite{Moore} and  in succession transform RBLMs to GAs, and
then determine the equivalence for GAs using the results obtained
on PAs. As indicated before, transforming  RBLMs to GAs is not
necessary when dealing with the equivalence between RBLMs.

Due to their complex  behaviors, CL-1QFAs and MM-1QFAs may not be
bilinearized as  Moore and  Crutchfield \cite{Moore} did for
MO-1QFAs. Hence, we need new ways to deal with them. Indeed, we
find that Bertoni et al \cite{Ber03} provided a useful technique
to our problem. In the following two subsections, we will focus on
determining the equivalence for CL-1QFAs and MM-1QFAs.

\subsection {Determining the equivalence for CL-1QFAs}

Determining whether two CL-1QFAs ${\cal A}_1$ and ${\cal A}_2$ are
equivalent is to verify whether $f_{{\cal A}_1}(x\$)=f_{{\cal
A}_2}(x\$)$ holds for any $x\in\Sigma^{*}$.  We may learn
something from how we deal with the equivalence problems for
MO-1QFAs \cite{Bro,Kos} and QSMs \cite{Li}, where the ways have a
common point, that is, to firstly transform quantum machines to be
in a bilinear form and then use some knowledge on matrix and
linear space to deal with that. However, we can see that the
behavior of CL-1QFAs is more complex than those of MO-1QFAs and
QSMs. Then there may need some more elaborate work on them.

  Below,  we will give a key lemma  that allows us to transform CL-1QFAs
  to be in the bilinear form---RBLMs. Then,  the equivalence
  problem of CL-1QFAs is transformed to that of RBLMs which can be solved by using Proposition 5.
The  idea behind the following lemma mainly derives   from Ref.~
\cite{Ber03}.
\begin{Lm}
Any  $m$-state CL-1QFA ${\cal M}$ over the working alphabet
$\Gamma=\Sigma\cup\{\$\}$ with control language ${\cal L}$ can be
simulated by a $(km^2)$-state RBLM $\hat{{\cal M}}$ over $\Gamma$,
where factor $k$ is the number of states in the minimal DFA that
recognizes the control language ${\cal L}$.
\end{Lm}
\noindent\textbf{Proof.} Suppose that we have a CL-1QFA ${\cal
  M}=(Q, \pi,\{U(\sigma)\}_{\sigma\in\Gamma},{\cal O},{\cal L})$
with $m$ states, where observable ${\cal O}$ has eigenvalue set
${\cal C}$ and  projector set $\{P(c):c\in{\cal C}\}$. Since the
control language ${\cal L}\subseteq
  {\cal C}^{*}$ is regular, there exists a minimal DFA recognizing ${\cal L}$.
  Then we suppose that  DFA ${\cal A}=(S,\rho,\{M(c)\}_{c\in{\cal
  C}},\xi)$ recognizes ${\cal L}$ with $|S|=k$. Now we construct a RBLM $\hat{{\cal
  M}}=(\hat{S},\hat{\pi},\{\hat{
  M}(\sigma)\}_{\sigma\in\Gamma},\hat{\eta})$ as follows:
\begin{itemize}
\item $\hat{\pi}=(\pi\otimes\pi^*\otimes\rho)$, where the symbol
$*$ denotes conjugate operation;

\item  $\hat{M}(\sigma)=\Big(U(\sigma)\otimes U^*(\sigma)\otimes
I\Big).\Big(\sum_{c\in{\cal C}}P(c)\otimes P(c)\otimes M(c)\Big
)$;

\item  $\hat{\eta}=\sum^m_{k=1}e_k\otimes e_k\otimes\xi$, where
$e_k$ is the column vector having 1 only at the $k$th component
and 0s else.
\end{itemize}
  Then we have (denoting $\$$ by $x_{n+1}$):
\begin{align*}
&f_{\hat{{\cal
  M}}}(x_1\dots x_n
  \$)
  =\hat{\pi}\hat{M}(x_1)\dots\hat{M}(x_n)\hat{M}(\$)\hat{\eta}\\
  =&(\pi\otimes\pi^*\otimes\rho)\prod^{n+1}_{i=1}\Big(\big(U(x_i)\otimes
U^*(x_i)\otimes I\big).\big(\sum_{c\in{\cal C}}P(c)\otimes
P(c)\otimes M(c)\big )\Big).\big(\sum^m_{k=1}e_k\otimes
e_k\otimes\xi\big)\\
=&(\pi\otimes\pi^*\otimes\rho)\sum_{y=y_1\dots y_{n+1}\in{\cal
C}^{n+1}}\Big(\prod^{n+1}_{i=1}U(x_i)P(y_i)\otimes
\prod^{n+1}_{i=1}U^{*}(x_i)P(y_i)\otimes\prod^{n+1}_{i=1}M(y_i)\Big).\big(\sum^m_{k=1}e_k\otimes
e_k\otimes\xi\big)\\
 =&\sum^m_{k=1}\sum_{y=y_1\dots y_{n+1}\in{\cal
C}^{n+1}}\Big(\pi\prod^{n+1}_{i=1}U(x_i)P(y_i)\Big)_k\Big(\pi^*\prod^{n+1}_{i=1}U^*(x_i)P(y_i)\Big)_k\rho
M(y)\xi\\
=&\sum_{y=y_1\dots y_{n+1}\in{\cal C}^{n+1}}{\cal X}_{\cal
L}(y)\sum^m_{k=1}|(\pi\prod^{n+1}_{i=1}U(x_i)P(y_i))_k|^2\\
=&\sum_{y=y_1\dots y_{n+1}\in{\cal
L}}||\pi\prod^{n+1}_{i=1}U(x_i)P(y_i)||^2\\
=&f_{\cal M}(x_1\dots x_n\$).
\end{align*}

We have shown that ${\cal M}$ and $\hat{\cal M}$ have the same
behavior for any word $w\in\Sigma^*\$$, and  $\hat{\cal M}$ has
$km^2$ states.
\qed\\
\begin{Rm}
One can find that in the above process, the DFA recognizing the
control language ${\cal L}$ has no need to be necessarily minimal.
In practice, when some DFA recognizing the control language is
given, we can construct the RBLM by using it. However, as we can
see, the minimal DFA can keep the resulted RBLM as small as
possible, and then leads to a  tight bound in Theorem 7 as
follows.
\end{Rm}

\begin{Th}
Two CL-1QFAs ${\cal A}_1$ and ${\cal A}_2$ with control languages
${\cal L}_1$ and  ${\cal L}_2$, respectively, are equivalent if
and only if they are $(c_1n_1^2+c_2n_2^2-1)$-equivalent, where
$n_1$ and $n_2$ are the numbers of states in ${\cal A}_1$ and
${\cal A}_2$, respectively, and $c_1$ and $c_2$ are the numbers of
states in the minimal DFAs that recognize ${\cal L}_1$ and  ${\cal
L}_2$, respectively. Furthermore, if ${\cal L}_1$ and ${\cal L}_2$
are given in the form of DFAs, with $m_1$ and $m_2$ states,
respectively, then there exists a polynomial-time algorithm
running in time $O\big((m_1n_1^2+m_2n_2^2)^4\big)$ that takes as
input ${\cal A}_1$ and ${\cal A}_2$ and determines whether they
are equivalent.
\end{Th}
\noindent\textbf{Proof.} Suppose that CL-1QFAs ${\cal A}_1$ and
${\cal A}_2$ with control languages ${\cal L}_1$ and ${\cal L}_2$,
respectively, have the same input alphabet $\Sigma$ and the
end-marker $\$$, and that ${\cal L}_1$ and  ${\cal L}_2$ can be
recognized by the minimal DFAs with $c_1$ and $c_2$ states,
respectively. Now we have to determine whether $f_{{\cal
A}_1}(w\$)=f_{{\cal A}_2}(w\$)$ holds for any $w\in\Sigma^*$. We
do that in  the following steps, where we firstly transform
CL-1QFAs to RBLMs, then remove the end-maker $\$$, and lastly
determine the equivalence for RBLMs.
\begin{enumerate}
     \item [(1)] By
Lemma 6, ${\cal A}_1$ and ${\cal A}_2$ can be simulated by  two
RBLMs ${\cal A}^{(1)}_1$ and ${\cal A}^{(1)}_2$ over the alphabet
$\Sigma\cup\{\$\}$ with $c_1n_1^2$ and $c_2n_2^2$ states,
respectively, such that $f_{{\cal A}_1}(w\$)=f_{{\cal
A}^{(1)}_1}(w\$)$ and $f_{{\cal A}_2}(w\$)=f_{{\cal
A}^{(1)}_2}(w\$)$ for any $w\in \Sigma^{*}$.

 \item [(2)] By Proposition 1, there are two RBLMs ${\cal A}^{(2)}_1$ and ${\cal
 A}^{(2)}_2$ over the alphabet $\Sigma$, with $c_1n_1^2$ and $c_2n_2^2$ states,
respectively, such that $f_{{\cal A}^{(1)}_1}(w\$)=f_{{\cal
A}^{(2)}_1}(w)$ and $f_{{\cal A}^{(1)}_2}(w\$)=f_{{\cal
A}^{(2)}_2}(w)$.

 \item [(3)]  By Definition 2 and Proposition 5,
 $f_{{\cal A}^{(2)}_1}(w)=f_{{\cal A}^{(2)}_2}(w)$ holds for any
 $w\in\Sigma^*$ iff it holds for any $w\in\Sigma^*$ with $|w|\leq
 c_1n_1^2+c_2n_2^2-1$.
\end{enumerate}

Therefore, $f_{{\cal A}_1}(w\$)=f_{{\cal A}_2}(w\$)$ holds for any
$w\in\Sigma^*$ if and only if it holds for any $w\in\Sigma^*$ with
$|w|\leq
 c_1n_1^2+c_2n_2^2-1$.

Furthermore,  if we want to design an algorithm that simulates the
above steps  to determine whether ${\cal A}_1$ and ${\cal A}_2$
are equivalent, then the consuming time will vary with the given
forms of ${\cal L}_1$ and ${\cal L}_2$:
\begin{enumerate}
\item[(i)]
  ${\cal L}_1$ and ${\cal L}_2$ are given in the form of
regular expressions. Then, according to the results in \cite{Hop},
it will need exponential time (in the lengths of ${\cal L}_1$ and
${\cal L}_2$) to construct DFAs from ${\cal L}_1$ and ${\cal L}_2$
in step (1), and as a result, the  total  time will have an
exponential additive factor.
 \item[(ii)]
  ${\cal L}_1$ and ${\cal L}_2$ are given in the
form of DFAs (not necessarily in minimal form), say $M_1$ and
$M_2$ with $m_1$ and $m_2$ states, respectively. Recall that we
have assumed that the operations of addition and multiplication on
 two complex numbers can all be done in constant time. Then, from the proof of Lemma 6, we can find that step
(1) consumes time $O\big((m_1 n_1^2)^3+(m_2 n_2^2)^3\big)$  that
is mainly used on the multiplication and Kronecker product of
matrices, producing two RBLMs ${\cal A}^{(1)}_1$ and ${\cal
A}^{(1)}_2$ with $m_1n_1^2$ and $m_2n_2^2$ states, respectively.
Step (2) taking as input ${\cal A}^{(1)}_1$ and ${\cal A}^{(1)}_2$
can be done in time $O\big((m_1 n_1^2)^2+(m_2 n_2^2)^2\big)$. From
Proposition 5, step (3) taking as input two RBLMs with $m_1n_1^2$
and $m_2n_2^2$ states, respectively, can be done in time
$O\big((m_1n_1^2+m_2n_2^2)^4\big)$. Therefore, the total time is
$O\big((m_1n_1^2+m_2n_2^2)^4\big)$.
\end{enumerate}

  Now we have proven the theorem. \qed\\
\begin{Rm}
There may be a  better solution to the problem of determining the
equivalence between CL-1QFAs. Nevertheless, the current good news
is that Theorem 7 indeed provides a bound on the length of strings
that need to be verified when we want to determine the equivalence
between two CL-1QFAs.
\end{Rm}

\subsection {Determining the equivalence for MM-1QFAs}
Gruska \cite{Gru00} proposed as an open problem that is it
decidable whether two MM-1QFAs are equivalent. Then Koshiba
\cite{Kos} tried to solve the problem. His method consists of two
steps: (i) for any MM-1QFA,  construct an equivalent MO-g1QFA
(like MO-1QFA but with transformation matrices not necessarily
unitary); (ii) determine the equivalence for MO-g1QFAs using the
known way on MO-1QFAs. Nevertheless, we find that the construction
technique stated in \cite{Kos} for step (i) is not valid, i.e., it
produces an MO-g1QFA that is not equivalent to the original
MM-1QFA. Thus, the problem is in fact not solved there. Below, we
will give a detailed explanation of this invalidity.

\subsubsection{The invalidity of Koshiba's way}
Note that when we show the invalidity in the following, we will
adopt the definitions of QFAs stated in \cite{Bro} that have
slight difference from the definitions stated before.

 First
let us  recall the way stated in [20, Theorem 3] for constructing
MO-g1QFAs from MM-1QFAs. Given an MM-1QFA ${\cal
M}=(Q,\Sigma,\{U_\sigma\}_{\sigma\in\Sigma\cup\{\$\}},q_0,
Q_{acc},Q_{rej})$,   an MO-g1QFA ${\cal
M}^{'}=(Q^{'},\Sigma,\{U^{'}_\sigma\}_{\sigma\in\Sigma\cup\{\$\}},q_0,
F)$ is constructed as follows:
\begin{itemize}
    \item  $Q^{'}=Q\cup\{q_\sigma: \sigma\in\Sigma\cup\{\$\}\}\backslash Q_{acc}$, and $F=\{q_\sigma: \sigma\in\Sigma\cup\{\$\}\}$;
    \item  $ U^{'}_\sigma|q\rangle=\dots + \alpha_i|q_i\rangle\dots
    +\alpha_A|q_\sigma\rangle$ when $U_\sigma|q\rangle=\dots + \alpha_i|q_i\rangle\dots +\alpha_A|q_A\rangle$ and $q_A\in Q_{acc}$
  ;
    \item  add the rules:
    $U^{'}_{\sigma}|q_\sigma\rangle=|q_\sigma\rangle$ for all
    $|q_\sigma\rangle\in F$.
    \end{itemize}

Koshiba \cite{Kos} deemed that the construction technique stated
above can ensure that for any input word, the accepting
probability in ${\cal M}$  is preserved in ${\cal M}^{'}$, which
is in fact not so. Firstly, the transformation stated above is
unclear, since in the general case $|Q_{acc}|>1$, the second rule
is unclear. Secondly, even in the simplest case $|Q_{acc}|=1$, the
transformation is not valid. The essential reason for the
invalidity of the above way is that the accepting state set $F$ in
${\cal M}^{'}$ does not cumulate the accepting probabilities in
the original MM-1QFA. Instead, it accumulates just the accepting
amplitudes. In addition, we know that in general,
$|a|^2+|b|^2\neq|a+b|^2$. Therefore, the above way  leads to
invalidity. For concreteness, we provide a counterexample to show
the invalidity for the case $|Q_{acc}|=1$ below.
 \paragraph{ A counterexample} Let MM-1QFA ${\cal
M}=(Q,\Sigma,\{U_\sigma\}_{\sigma\in\Sigma\cup\{\$\}},q_0,
Q_{acc},Q_{rej})$, where $Q=\{q_0,q_1,q_{acc},q_{rej}\}$ with the
set of accepting states $Q_{acc}=\{q_{acc}\}$ and the set of
rejecting states $Q_{rej}=\{q_{rej}\}$; $\Sigma=\{a\}$; $q_0$ is
the initial state; $\{U_\sigma\}_{\sigma\in\Sigma\cup\{\$\}}$ are
described below.
\begin{align*}
&U_a(|q_0\rangle)=
\frac{1}{2}|q_0\rangle+\frac{1}{\sqrt{2}}|q_1\rangle+\frac{1}{2}|q_{acc}\rangle,\\
&U_a(|q_1\rangle)=
\frac{1}{2}|q_0\rangle-\frac{1}{\sqrt{2}}|q_1\rangle+\frac{1}{2}|q_{acc}\rangle,\\
&U_\$(|q_0\rangle)=|q_{acc}\rangle,\hspace{2mm}U_\$(|q_1\rangle)=|q_{rej}\rangle.
\end{align*}

Next, we show how this automaton works on the input word $aa\$$.
\begin{enumerate}
\item The automaton starts in $|q_0\rangle$. Then $U_a$ is
applied, giving
$\frac{1}{2}|q_0\rangle+\frac{1}{\sqrt{2}}|q_1\rangle+\frac{1}{2}|q_{acc}\rangle$.
This state is measured with two possible outcomes produced. With
probability $(\frac{1}{2})^2$, the accepting state is observed,
and then the computation terminates. Otherwise, a non-halting
state $\frac{1}{2}|q_0\rangle+\frac{1}{\sqrt{2}}|q_1\rangle$
(unnormalized) is observed,
 and then the computation continues.

\item After the second $a$ is fed, the state
$\frac{1}{2}|q_0\rangle+\frac{1}{\sqrt{2}}|q_1\rangle$ is mapped
to
$\frac{1}{2}(\frac{1}{2}+\frac{1}{\sqrt{2}})|q_0\rangle+\frac{1}{\sqrt{2}}(\frac{1}{2}-\frac{1}{\sqrt{2}})
|q_1\rangle+\frac{1}{2}(\frac{1}{2}+\frac{1}{\sqrt{2}})|q_{acc}\rangle$.
This is measured with two possible outcomes. With probability
$[\frac{1}{2}(\frac{1}{2}+\frac{1}{\sqrt{2}})]^2$, the computation
terminates in the accepting state $q_{acc}$. Otherwise, the
computation continues with a new no-halting state
$\frac{1}{2}(\frac{1}{2}+\frac{1}{\sqrt{2}})|q_0\rangle+\frac{1}{\sqrt{2}}(\frac{1}{2}-\frac{1}{\sqrt{2}})
|q_1\rangle$ (unnormalized).

\item After the last symbol $\$$ is fed, the automaton's state
turns to
$\frac{1}{2}(\frac{1}{2}+\frac{1}{\sqrt{2}})|q_{acc}\rangle+\frac{1}{\sqrt{2}}(\frac{1}{2}-\frac{1}{\sqrt{2}})
|q_{rej}\rangle$. This is measured. The computation terminates in
the accepting state $|q_{acc}\rangle$ with probability
$[\frac{1}{2}(\frac{1}{2}+\frac{1}{\sqrt{2}})]^2$ or in the
rejecting state $|q_{rej}\rangle$ with probability
$[\frac{1}{\sqrt{2}}(\frac{1}{2}-\frac{1}{\sqrt{2}})]^2$.
\end{enumerate}

 The total accepting probability  is
$(\frac{1}{2})^2+[\frac{1}{2}(\frac{1}{2}+\frac{1}{\sqrt{2}})]^2+[\frac{1}{2}(\frac{1}{2}+\frac{1}{\sqrt{2}})]^2=
\frac{5}{8}+\frac{1}{2\sqrt{2}}$.

Note that in the above steps, we did not normalize the
intermediate states produced. As we know, according to quantum
mechanics, after every measurement, the states should be
normalized. However, in the above process, adopting the
unnormalized states makes the representation of states simple and
the calculation of accepting probability convenient, and still
keeps the correctness of the total accepting probability. This
strategy was also used by Ambainis and Freivalds \cite{Amb-F}.

 Now according to the construction technique [19,
Theorem 3] stated before, we get an MO-g1QFA ${\cal
M}^{'}=(Q^{'},\Sigma,\{U_\sigma^{'}\}_{\sigma\in\Sigma\cup\{\$\}},q_0,F)$
where $Q^{'}=\{q_0,q_1, q_{rej}, q_a, q_\$\}$, $F=\{q_a, q_\$\}$
and $\{U_\sigma^{'}\}_{\sigma\in\Sigma\cup\{\$\}}$ are described
below.
\begin{align*}
&U^{'}_a(|q_0\rangle)=
\frac{1}{2}|q_0\rangle+\frac{1}{\sqrt{2}}|q_1\rangle+\frac{1}{2}|q_a\rangle,\\
&U^{'}_a(|q_1\rangle)=
\frac{1}{2}|q_0\rangle-\frac{1}{\sqrt{2}}|q_1\rangle+\frac{1}{2}|q_a\rangle,\\
&U^{'}_\$(|q_0\rangle)=|q_\$\rangle,\hspace{2mm}U^{'}_\$(|q_1\rangle)=|q_{rej}\rangle,\\
&U^{'}_a(|q_a\rangle)=|q_a\rangle,\hspace{2mm}U^{'}_\$(|q_a\rangle)=|q_a\rangle.
\end{align*}

When the input word is $aa\$$, the automaton works as follows.
Starting from state $|q_0\rangle$, when the first $a$ is fed, the
automaton turns to state
$\frac{1}{2}|q_0\rangle+\frac{1}{\sqrt{2}}|q_1\rangle+\frac{1}{2}|q_a\rangle$.
After the second $a$ is fed, the state is mapped to
$\frac{1}{2}(\frac{1}{2}+\frac{1}{\sqrt{2}})|q_0\rangle+\frac{1}{\sqrt{2}}(\frac{1}{2}-\frac{1}{\sqrt{2}})
|q_1\rangle+[\frac{1}{2}+\frac{1}{2}(\frac{1}{2}+\frac{1}{\sqrt{2}})]|q_a\rangle$.
After the last symbol $\$$ is fed, the state is mapped to
$\frac{1}{2}(\frac{1}{2}+\frac{1}{\sqrt{2}})|q_\$\rangle+\frac{1}{\sqrt{2}}(\frac{1}{2}-\frac{1}{\sqrt{2}})
|q_{rej}\rangle+[\frac{1}{2}+\frac{1}{2}(\frac{1}{2}+\frac{1}{\sqrt{2}})]|q_a\rangle$.

The total accepting probability is
$[\frac{1}{2}(\frac{1}{2}+\frac{1}{\sqrt{2}})]^2+[\frac{1}{2}+\frac{1}{2}(\frac{1}{2}+\frac{1}{\sqrt{2}})]^2=\frac{7}{8}+\frac{1}{\sqrt{2}}$.

Now it turns out  that the accepting probability in the original
MM-1QFA is not preserved in the constructed machine as expected in
\cite{Kos}. Therefore, the invalidity of the method of [20,
Theorem 3] has been shown.

\subsubsection{Our way for deciding the equivalence between MM-1QFAs }
  As stated before, due to the complex  behavior of MM-1QFAs, it is
likely no longer valid to deal with MM-1QFAs as Moore and
Crutchfield \cite{Moore} did for MO-1QFAs. At the same time, we
have shown    that  Koshiba's method \cite{Kos} is not valid to
decide whether two MM-1QFAs are equivalent. In addition, to our
knowledge, so far there seems to  have been no existing valid
solution to this problem. Therefore,  we would like to do that in
the following.

Now we  try to determine the equivalence between MM-1QFAs,
starting by a proposition  introduced as follows.
\begin{Pp}[\cite{Ber03}]
Let $U(\sigma)$ be a unitary matrix, for $\sigma\in \Sigma$, and
${\cal O}$ an observable with results in ${\cal C}$, described by
projectors $P(c)$, for $c\in{\cal C}$. For any complex vector
$\alpha$ and any word $x=x_1\dots x_r\in\Sigma^r$, we get
\begin{align*}
\sum_{y_1\dots y_r\in {\cal C}^r}||\alpha\prod_{i=1}^r
U(x_i)P(y_i)||^2=||\alpha||^2.
\end{align*}
\end{Pp}
\noindent\textbf{Proof.} Using the properties of unitary matrices
and projective measurement, it is easy to prove  this proposition
by induction on the length of $x$. \qed\\

  Based on \cite{Ber03}, we get another key
lemma. With this lemma, we can transform MM-1QFAs to  CL-1QFAs for
which the equivalence problem  has been solved.

\begin{Lm}
Given an MM-1QFA ${\cal
M}=(Q,\pi,\{U(\sigma)\}_{\sigma\in\Sigma\cup\{\$\}},{\cal O})$,
there is a CL-1QFA ${\cal
M}^{'}=(Q,\pi,\{U(\sigma)\}_{\sigma\in\Sigma\cup\{\$\}},{\cal
O},g^*a\{a,r,g\}^*)$ such that for any $w\in\Sigma^*$, $f_{\cal
M}(w\$)=f_{{\cal M}^{'}}(w\$)$.
\end{Lm}
\noindent\textbf{Proof.}  Suppose that there are MM-1QFA ${\cal
M}$ and CL-1QFA ${\cal M}^{'}$ as stated above. For any $x_1\dots
x_n\in \Sigma^*$, there is (denoting $\$$ by $x_{n+1}$):
\begin{align*}
&f_{{\cal M}^{'}}(x_1\dots x_n\$)\\ =&\sum_{y_1\dots y_{n+1}\in
g^*a\{a,r,g\}^*}||\pi\prod^{n+1}_{i=1}U(x_i)P(y_i)||^2\\
=&\sum^n_{k=0}\sum_{y_{k+2}\dots
y_{n+1}}||\pi\prod^k_{i=1}\big(U(x_i)P(g)\big)U(x_{k+1})P(a)\prod^{n+1}_{j=k+2}U(x_j)P(y_j)||^2\\
=&\sum^n_{k=0}||\pi\prod^k_{i=1}\big(U(x_i)P(g)\big)U(x_{k+1})P(a)||^2\hspace{2mm}(\text{by Proposition 8})\\
=&f_{{\cal M}}(x_1\dots x_n\$)\hspace{2mm}(\text{by Eq. (4)}).
\end{align*}
Note that the two automata have the same states. We end the proof
here.
\qed\\

Now we  obtain the following theorem that determines the
equivalence between two MM-1QFAs.
\begin{Th}
Two MM-1QFAs ${\cal A}_1$ and ${\cal A}_2$  with $n_1$ and $n_2$
states, respectively, are equivalent if and only if they are
$(3n_1^2+3n_2^2-1)$-equivalent. Furthermore, there is a
polynomial-time algorithm running in time
$O\big((3n_1^2+3n_2^2)^4\big)$ that takes as input ${\cal A}_1$
and ${\cal A}_2$
 and determines whether ${\cal A}_1$ and ${\cal A}_2$ are
equivalent.
\end{Th}
\noindent\textbf{Proof.} Suppose that MM-1QFAs ${\cal A}_1$ and
${\cal A}_2$ with $n_1$ and $n_2$ states, respectively, have the
same input alphabet $\Sigma$ and the end-marker $\$$. Now we
determine whether $f_{{\cal A}_1}(w\$)=f_{{\cal A}_2}(w\$)$ holds
for any $w\in\Sigma^*$. We can do that by the following steps.
\begin{enumerate}
 \item[(1)]By
Lemma 9, ${\cal A}_1$ and ${\cal A}_2$ can be transformed into two
CL-1QFAs ${\cal A}^{(1)}_1$ and ${\cal A}^{(1)}_2$ over the
working alphabet $\Gamma=\Sigma\cup\{\$\}$ with $n_1$ and $n_2$
states, respectively, both of which have the same constant control
language $g^*a\{a,r,g\}^*$.

 \item[(2)] By Lemma 6, ${\cal A}^{(1)}_1$ and ${\cal
A}^{(1)}_2$ can be transformed into two RBLMs ${\cal A}^{(2)}_1$
and ${\cal A}^{(2)}_2$ over $\Gamma$, with $3n_1^2$ and $3n_2^2$
states, respectively, where the factor $3$ is the number of states
in the DFA (described in Fig.~1) recognizing the control language
$g^*a\{a,r,g\}^*$.

\item[(3)]  By Proposition 1,  we can construct ${\cal A}^{(3)}_1$
and ${\cal A}^{(3)}_2$ over the alphabet $\Sigma$ from ${\cal
A}^{(2)}_1$ and ${\cal A}^{(2)}_2$, such that $f_{{\cal
A}_1}(w\$)=f_{{\cal A}^{(3)}_1}(w)$ and  $f_{{\cal
A}_2}(w\$)=f_{{\cal A}^{(3)}_2}(w)$ for any $w\in\Sigma^*$.
Therefore, determining whether $f_{{\cal A}_1}(w\$)=f_{{\cal
A}_2}(w\$)$ holds for any $w\in\Sigma^*$ is equivalent to
determining whether ${\cal A}^{(3)}_1$ and ${\cal A}^{(3)}_2$ are
equivalent.

\item [(4)] By Proposition 5, ${\cal A}^{(3)}_1$ and ${\cal
A}^{(3)}_2$ are equivalent if and only if they are
$(3n_1^2+3n_2^2-1)$-equivalent.
\end{enumerate}

Therefore, $f_{{\cal A}_1}(w\$)=f_{{\cal A}_2}(w\$)$  holds for
any $w\in\Sigma^*$ if and only if it holds for any $w\in\Sigma^*$
with $|w|\leq
 3n_1^2+3n_2^2-1$.
Furthermore,  It is readily seen that step (1)  can be done in
constant time, and the other steps can be done in time
$O\big((3n_1^2+3n_2^2)^4\big)$ from the proof of Theorem 7.
Therefore, there exits a polynomial-time algorithm simulating the
above steps to determine whether two MM-1QFAs are equivalent.
Hence, we have completed the proof.\qed\\
\begin{center}
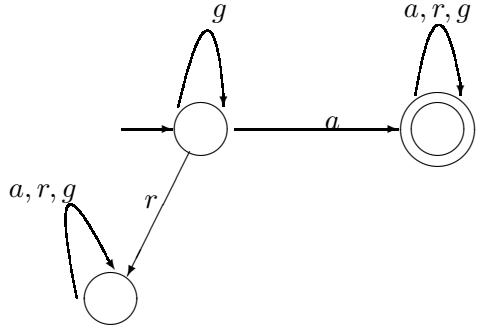
{\small Fig 1. The DFA  recognizing regular language $g^*a\{a,r,g\}^*$}\\
 \setlength{\unitlength}{1.5mm}
 \begin{picture}(35,32)
\put(7,20){\vector(1,0){4.5}} \put(14,20){\circle{5}} \qbezier
(12,22) (15,35) (16,23)
\put(16,23){\vector(0,-1){1}}\put(15,30){$g$}

\put(17,20){\vector(1,0){14.5}} \put(25,20){$a$}

 \put(35,20){\circle{5}}
\put(35,20){\circle{7}} \qbezier(33,23) (35,35) (37,24)
\put(37,24){\vector(0,-1){1}} \put(32,30){$a,r,g$}

\put(13,18){\vector(-1,-2){5.5}} \put(9,13){$r$}

 \put(6,5){\circle{5}}
\qbezier(3,5) (0,20) (6,8)
\put(5.4,9){\vector(2,-3){1}}\put(-3,14){$a,r,g$}
 \end{picture}
\end{center}

\section{Conclusions}
 QFAs are simple but basic models of quantum computation, but the decidability problem
for equivalence between QFAs has not been solved completely. In
this paper, we considered the decision of equivalence for CL-1QFAs
and MM-1QFAs. Specifically, we have shown that two CL-1QFAs ${\cal
A}_1$ and ${\cal A}_2$ with control languages (regular languages)
${\cal L}_1$ and  ${\cal L}_2$, respectively, are equivalent if
and only if they are $(c_1n_1^2+c_2n_2^2-1)$-equivalent, where
$n_1$ and $n_2$ are the numbers of states in ${\cal A}_1$ and
${\cal A}_2$, respectively, and $c_1$ and $c_2$ are the numbers of
states in the minimal DFAs that recognize ${\cal L}_1$ and  ${\cal
L}_2$, respectively. Furthermore, given ${\cal L}_1$ and ${\cal
L}_2$ in the form of DFAs, with $m_1$ and $m_2$ states,
respectively,  a polynomial-time algorithm was given,  that
determines whether ${\cal A}_1$ and ${\cal A}_2$ are equivalent in
time $O ((m_1n_1^2+m_2n_2^2)^4)$.

On the other hand, we clarified the existing error of the method
for determining the equivalence between  MM-1QFAs in the
literature \cite{Kos}. In particular, we showed that two MM-1QFAs
${\cal A}_1$ and ${\cal A}_2$ with $n_1$ and $n_2$ states,
respectively, are equivalent if, and only if they are
$(3n_1^2+3n_2^2-1)$-equivalent. Also, a polynomial-time algorithm
was presented, that determines whether ${\cal A}_1$ and ${\cal
A}_2$ are equivalent in time $O ((3n_1^2+3n_2^2)^4)$. Thus, the
problem proposed by Gruska \cite{Gru00} has been addressed.

So far, the equivalence issues for MO-1QFAs, MM-1QFAs, and
CL-1QFAs have been addressed. However, the equivalence concerning
another important model---2QFAs \cite{Kon97} is still open and
worthy of further consideration.

\end{document}